\documentstyle[12pt]{article}
\newcommand{\be}{\begin{equation}}
\newcommand{\ee}{\end{equation}}
\newcommand{\bea}{\begin{eqnarray}}
\newcommand{\eea}{\end{eqnarray}}
\newcommand{\bd}{\begin{displaymath}}
\newcommand{\ed}{\end{displaymath}}

\newcommand{\half}{\frac{1}{2}}
\newcommand{\lnb}{\mbox{\Large$\left(\right.$}}
\newcommand{\rnb}{\mbox{\Large$\left.\right)$}}
\newcommand{\lsb}{\mbox{\LARGE$\left[\right.$}}
\newcommand{\rsb}{\mbox{\LARGE$\left.\right]$}}

\newcommand{\LCB}{\mbox{\LARGE$\left\{\right.$}}
\newcommand{\RCB}{\mbox{\LARGE$\left.\right\}$}}

\begin{document}

\title{ Anomalous fermion number violation      \\
        and numerical simulations }

\author{ I. Montvay                             \\
 Deutsches Elektronen-Synchrotron DESY,         \\
 Notkestr. 85, D-22603 Hamburg, FRG }

\date{}
\maketitle

\begin{abstract}
 After discussing the problem of lattice regularization of chiral gauge
 theories, a simple model for anomalous fermion number violation
 is formulated which can be numerically studied with present day
 technique.
 Exploratory results of numerical simulations of a two-dimensional
 U(1) Higgs model are presented.
\end{abstract}

\section{ Introduction }

 The anomalous baryon and lepton number violation is a consequence of
 the chirality of the electroweak gauge interactions \cite{THOOFT}.
 This is a non-perturbative phenomenon and, therefore, a lattice
 formulation allowing for numerical simulations would be desirable.
 It is, however, well known that the non-perturbative formulation of
 chiral gauge theories is problematic
 (see, for instance, \cite{ROMAPR}).

 In the first part of my talk some recent developments concerning
 lattice chiral fermions are reviewed and the possibility of
 existence of mirror fermions is discussed.
 Recent work on the limits of mirror fermion parameters is reviewed
 within the framework of a particular mixing scheme of
 fermion--mirror-fermion pairs.
 In the second part a simple model of anomalous fermion number
 violation is discussed which can be studied on the
 lattice \cite{ANFEVI}.
 In the last part a simplified toy model in two dimensions, with
 chiral U(1)-symmetry, is discussed and some recent numerical
 results concerning topological charge and Chern-Simons number
 are presented.

\section{ Chiral fermions on the lattice? }

\subsection{ Recent developments }

 For an exact lattice formulation of the Standard Model on the
 lattice one has to solve the problem of chiral gauge theories.
 The recently proposed {\em ``domain wall fermion''} formulation
 \cite{CALHAR,KAPLAN} can potentially lead to a solution by
 introducing an additional fifth dimension with
 four dimensional hypersurfaces representing the real world.
 A simpler, more economic, version of this idea is to throw away
 half of the five dimensional space, resulting in the
 {\em ``boundary fermion''} formulation \cite{SHAMIR1}.
 The difficulty in these formulations arises at switching on
 the gauge coupling.
 If in the fifth direction the gauge coupling is chosen to be
 strong, one has to fight against the dangerous {\em ``layered phase''}
 \cite{FUNIEL}, where subsequent four-dimensional hyperplanes do
 not interact and the theory becomes vector-like \cite{KONIPR}.
 In case of a four-dimensional gauge field, identical in all
 hyperplanes, the decoupling of the chiral fermion from its
 mirror partner with opposite chirality seems unlikely, at least as
 long as exact gauge invariance is maintained.

 A further obstacle for defining chiral gauge theories on the
 lattice is a recent extension of the Nielsen-Ninomiya no-go
 theorem including interactions and considering the continuum
 limit \cite{SHAMIR2}.

 Avoiding the fermion doubling problem on random lattices has
 recently been also investigated, and does not seem to be promising,
 either \cite{GRIKIE}.

 The remaining possibilities for the lattice formulation of chiral
 gauge theories are not explicitly gauge invariant.
 In the modified Rome-approach capable to accomodate also fermion
 number violations \cite{ROMA2} gauge fixing is essential.
 In case of the {\em ``reduced staggered fermions''} \cite{SMITST}
 gauge symmetry is explicitly broken by lattice actifacts, but
 hopefully restored by the dynamics.

\subsection{ Mirror fermions }

 Since the non-perturbative formulation of the electroweak
 interactions is so difficult, a natural question is whether
 ``chirality'' is perhaps only a low-energy phenomenon, and at
 high energy the space-reflection symmetry is restored by the
 existence of opposite chirality {\em ``mirror fer\-mi\-ons''}
 \cite{LEEYAN}.
 If the presently known (almost complete) three fermion families
 were duplicated at the electroweak energy scale, in the range
 100-1000 GeV, by three mirror fermion families with opposite
 chiralities and hence V+A couplings to the weak gauge vector
 bosons \cite{MIRFAM}, then the whole fermion spectrum would be
 ``vectorlike''.
 This would very much simplify the non-perturbative lattice formulation
 of the Standard Model \cite{MIRFER}.

 The direct pair production of mirror fermions is not observed at LEP.
 This puts a lower limit on their masses of about 45 GeV.
 Heavier mirror fermions could be produced via their mixing to
 ordinary fermions.
 The present data imply some constraints on the mixing angles versus
 the masses which, however, strongly depend on the mixing scheme.
 (For an evaluation of the constraints implied by the high
 precision LEP data see the recent paper by Cs\'aki and Csikor
 \cite{CSACSI}).

\subsection{ Fermion--mirror-fermion mixing scheme }

 The strongest constraints on mixing angles between ordinary
 fermions and mirror fermions arise from the
 conservation of $e$-, $\mu$- and $\tau$- lepton numbers and from
 the absence of flavour changing neutral hadronic currents.
 In a particular scheme these constraints can be implemented at tree
 level \cite{MIRFAM}.
 In this {\em ``monogamous mixing''} scheme the structure of the
 mass matrix is such that there is a one-to-one correspondence between
 fermions and mirror fermions, due to the fact that the family
 structure of the mass matrix for mirror fermions is closely related
 to the one for ordinary fermions.

 Let us denote doublet indices by $A=1,2$, colour indices by
 $c=1,2,3$ in such a way that the leptons belong to the fourth value of
 colour $c=4$, and family indices by $K=1,2,3$.
 Diagonal entries in the mass matrix for ``normal'' fermions will
 be denoted by an index $\psi$, those for the mirror fermions
 by $\chi$ and the off-diagonal mixing masses between fermions and
 mirror fermions by $L$ or $R$ (depending on the chirality of the
 ``normal'' fermion).
 In general the elements of the mass matrix for three mirror pairs
 of fermion families are diagonal in isospin and colour, hence they
 have the form
$$
\mu_{(\psi,\chi);A_2c_2K_2,A_1c_1K_1} = \delta_{A_2A_1}
\delta_{c_2c_1} \mu^{(A_1c_1)}_{(\psi,\chi);K_2K_1}  \ ,
$$
\be \label{eq01}
\mu_{L;A_2c_2K_2,A_1c_1K_1} = \delta_{A_2A_1}
\delta_{c_2c_1} \mu^{(c_1)}_{L;K_2K_1}  \ ,
\hspace{3em}
\mu_{R;A_2c_2K_2,A_1c_1K_1} = \delta_{A_2A_1}
\delta_{c_2c_1} \mu^{(A_1c_1)}_{R;K_2K_1}  \ .
\ee
 The diagonalization of the mass matrix $M$ can be achieved for given
 indices $A$ and $c$ by two $6 \otimes 6$ unitary matrices $F_L^{(Ac)}$
 and $F_R^{(Ac)}$ acting, respectively, on the L-handed and R-handed
 subspaces:
\be \label{eq02}
F_L^{(Ac)\dagger}(M^\dagger M)_L F_L^{(Ac)} \ ,
\hspace{3em}
F_R^{(Ac)\dagger}(M^\dagger M)_R F_R^{(Ac)} \ .
\ee

 The main assumption of the ``monogamous'' mixing scheme is that
 in the family space $\mu_\psi,\mu_\chi,\mu_L,\mu_R$ are hermitian
 and simultaneously diagonalizable, that is
\be \label{eq03}
F_L^{(Ac)} = F_R^{(Ac)} =  \left(
\begin{array}{cc}
F^{(Ac)}  &  0  \\  0  &  F^{(Ac)}
\end{array} \right)  \ ,
\ee
 where the block matrix acts in $(\psi,\chi)$-space.
 The Kobayashi-Maskawa matrix of quarks is given by
\be \label{eq04}
C \equiv F^{(2c)\dagger} F^{(1c)} \ ,
\ee
 independently for $c=1,2,3$.
 The corresponding matrix with $c=4$ and $A=1 \leftrightarrow 2$
 describes the mixing of neutrinos, if the Dirac-mass of the neutrinos
 is nonzero.
 (Majorana masses of the neutrinos are not considered here, but
 in principle, they can also be introduced.)

 An example for a mass matrix with ``monogamous'' mixing is the
 following:
$$
\mu^{(Ac)}_{\chi;K_2K_1} = \lambda^{(Ac)}_\chi
\mu^{(Ac)}_{\psi;K_2K_1} + \delta_{K_2K_1}\Delta^{(Ac)} \ ,
$$
\be \label{eq05}
\mu^{(c)}_{L;K_2K_1} = \delta_{K_2K_1}\delta^{(c)}_L \ ,
\hspace{3em}
\mu^{(Ac)}_{R;K_2K_1} = \lambda^{(Ac)}_R
\mu^{(Ac)}_{\psi;K_2K_1} + \delta_{K_2K_1}\delta^{(Ac)}_R \ ,
\ee
 where $\lambda_\chi^{(Ac)},\; \Delta^{(Ac)},\; \delta_L^{(c)},\;
 \lambda_R^{(Ac)},\; \delta_R^{(Ac)}$ do not depend on the family
 index.

 The full diagonalization of the mass matrix on the
 $(\psi_L,\psi_R,\chi_L,\chi_R)$ basis of all three family pairs is
 achieved by the $96 \otimes 96$ matrix
\be \label{eq06}
{\cal O}^{(LR)}_{A^\prime c^\prime K^\prime,AcK}
= \delta_{A^\prime A}\delta_{c^\prime c}
F^{(Ac)}_{K^\prime K}
\cdot \left(
\begin{array}{cc}
 \cos\alpha^{(AcK)}_L  &  0  \\
 0  &  \cos\alpha^{(AcK)}_R  \\
-\sin\alpha^{(AcK)}_L  &  0  \\
 0  & -\sin\alpha^{(AcK)}_R
\end{array}
\begin{array}{cc}
 \sin\alpha^{(AcK)}_L  &  0  \\
 0  &  \sin\alpha^{(AcK)}_R  \\
 \cos\alpha^{(AcK)}_L  &  0  \\
 0  &  \cos\alpha^{(AcK)}_R
\end{array}  \right) \ .
\ee
 $M^\dagger M$ is diagonalized by
${\cal O}^{(LR)\dagger} M^\dagger M {\cal O}^{(LR)},$
 and $M M^\dagger$ by
${\cal O}^{(RL)\dagger} M M^\dagger {\cal O}^{(RL)},$
 where ${\cal O}^{(RL)}$ is obtained from ${\cal O}^{(LR)}$ by
 $\alpha_L \leftrightarrow \alpha_R$.

 In case of $\mu_R=\mu_L$, which happens for instance in (\ref{eq05})
 if $\lambda_R=0$ and $\delta_R=\delta_L$, the left-handed and
 right-handed mixing angles are the same:
\be \label{eq07}
\alpha^{(AcK)} \equiv \alpha_L^{(AcK)} =
\alpha_R^{(AcK)} \ .
\ee
 In Ref.~\cite{MIRFAM} only this special case was considered.
 The importance of the left-right-asymmetric mixing was pointed out in
 Ref.~\cite{CSIFOD}, where the constraints arising from the measured
 values of anomalous magnetic moments were determined.
 It turned out that for the electron and muon the upper limit is
\be \label{eq08}
|\alpha_L \alpha_R| \le 0.0004 \ ,
\ee
 which is much stronger than the limits obtained from all other
 data \cite{LANLON}:
\be \label{eq09}
\alpha_L^2,\; \alpha_R^2 \le 0.02 \ .
\ee
 In case of the L-R asymmetric mixing the constraint (\ref{eq08})
 can be satisfied, for instance, if the right-handed mixing vanishes
 (or is very small): $\alpha_R \simeq 0$.

 The hypothetical mirror fermions can be discovered at the next
 generation of high energy colliders.
 At HERA the first family mirror fermions can be produced via mixing
 to ordinary fermions up to masses of about 200 GeV, if the mixing
 angles are close to their present upper limits \cite{CSIMON,HERA}.
 At SSC and LHC mirror lepton pair production can be observed up to
 masses of a few hundred GeV \cite{CSIKOR}.
 This has the advantage of being essentially independent of the small
 mixing.
 At a high energy $e^+e^-$ collider, e.~g. LEP-200 or NLC,
 mirror fermions can be pair produced and easily identified up to
 roughly half of the total energy, and also produced via mixing
 almost up to the total energy \cite{NLC}.

 In a model with fermion--mirror-fermion pairs the anomaly in the
 baryon- and lepton-number current is zero.
 Therefore in such models there is no anomalous fermion number
 violation at the elecroweak symmetry breaking scale.
 The baryon asymmetry of the Universe has to be produced at some higher
 energy scale, for instance, at the scale of grand unification
 \cite{GEOGLA}.

\section{ Fermion number anomaly on the lattice }

 If mirror fermions do not exist, the lattice formulation of the
 Standard Model is problematic and the non-perturbative effects in
 the electroweak sector, like the anomalous fermion number violation,
 cannot be studied by numerical simulations.
 There is, however, an approximation of the electroweak sector of the
 standard model which can be studied with standard lattice techniques,
 namely the limit when the
 $\rm SU(3)_{colour} \otimes U(1)_{hypercharge}$ gauge couplings are
 neglected \cite{ANFEVI}.

\subsection{ Model for fermion number violation }

 A simple prototype model is the standard $\rm SU(2)_L$ Higgs model
 coupled to an even number $2N_f$ of fermion doublets.
 In the standard model we have $N_f=6$ (for simplicity, we consider
 Dirac-neutrinos, but the massless neutral right-handed neutrinos
 decouple \cite{GOLPET}).
 One can take, for simplicity, $N_f=1$ but the extension to $N_f > 1$
 is trivial.
 The lattice action depends on the matrix scalar field
 $\varphi_x=\phi_{0x}+i\phi_{sx}\tau_s$ (with four real fields
 $\phi_{S=0,\ldots,3}$) and the fermion doublet fields $\psi_{(1,2)x}$:
\be \label{eq10}
S=S_{scalar}+S_{fermion} \ .
\ee
 The standard scalar-gauge Higgs-model action is
\be \label{eq11}
S_{scalar} = \frac{1}{4}\sum_x \left\{
m_0^2 {\rm Tr\,}(\varphi^\dagger_x\varphi_x)
+ \lambda \left[ {\rm Tr\,}(\varphi^\dagger_x\varphi_x) \right]^2
+ \sum_{\mu=\pm 1}^{\pm 4}
[ {\rm Tr\,}  (\varphi^\dagger_x \varphi_x)
- {\rm Tr\,}  (\varphi^\dagger_{x+\hat{\mu}}U_{x\mu}\varphi_x) ]
\right\} \ .
\ee
 The fermionic part contains the chiral gauge fields
 (with $U_{x\mu} \in \rm SU(2)$ and $P_{L,R}=(1 \mp \gamma_5)/2$)
\be \label{eq12}
U_{(L,R)x\mu}=P_{(L,R)} U_{x\mu}+P_{(R,L)} \ ,
\ee
 and is given by
\bd
S_{fermion} = \sum_x \LCB \frac{\mu_0}{2} \left[
  (\psi^T_{2x}\epsilon C \psi_{1x}) - (\psi^T_{1x}\epsilon C \psi_{2x})
+ (\overline{\psi}_{2x}\epsilon C \overline{\psi}^T_{1x})
- (\overline{\psi}_{1x}\epsilon C \overline{\psi}^T_{2x}) \right]
\ed
\bd
- \half \sum_\mu \lsb
  (\overline{\psi}_{1 x+\hat{\mu}} \gamma_\mu U_{Lx\mu} \psi_{1x})
+ (\overline{\psi}_{2 x+\hat{\mu}} \gamma_\mu U_{Lx\mu} \psi_{2x})
\ed
\bd
- \frac{r}{2} \lnb
  (\psi^T_{2x}\epsilon C \psi_{1x})
- (\psi^T_{2 x+\hat{\mu}}\epsilon C U_{Lx\mu} \psi_{1x})
- (\psi^T_{1x}\epsilon C \psi_{2x})
+ (\psi^T_{1 x+\hat{\mu}}\epsilon C U_{Lx\mu} \psi_{2x})
\ed
\bd
+ (\overline{\psi}_{2x}\epsilon C \overline{\psi}^T_{1x})
- (\overline{\psi}_{2 x+\hat{\mu}} U_{Rx\mu}
   \epsilon C \overline{\psi}^T_{1x})
- (\overline{\psi}_{1x}\epsilon C \overline{\psi}^T_{2x})
+ (\overline{\psi}_{1 x+\hat{\mu}} U_{Rx\mu}
  \epsilon C \overline{\psi}^T_{2x})
\rnb \rsb
\ed
\be \label{eq13}
+ (\overline{\psi}_{1Rx} G_1\varphi^+_x \psi_{1Lx})
+ (\overline{\psi}_{1Lx} \varphi_x G_1  \psi_{1Rx})
+ (\overline{\psi}_{2Rx} G_2\varphi^+_x \psi_{2Lx})
+ (\overline{\psi}_{2Lx} \varphi_x G_2  \psi_{2Rx}) \RCB \ .
\ee
 Here $\epsilon=i\tau_2$ acts in isospin space, and $C$ is the
 fermion charge conjugation matrix.
 The Yukawa-couplings $G_{1,2}$ can, in general, be arbitrary
 diagonal matrices in isospin space.
 In case of degenerate doublets $G_{1,2}$ are proportional to the unit
 matrix.

 Instead of the off-diagonal Majorana mass $\mu_0$ and Majorana-like
 Wilson term (proportional to $r$), it is technically more convenient
 to consider a Dirac-like form with $\psi \equiv \psi_1$ and the
 mirror fermion field $\chi$ defined by
\be \label{eq14}
\chi_x \equiv \epsilon^{-1} C \overline{\psi}_{2x}^T \ ,
\hspace{2em}
\overline{\chi}_x \equiv \psi_{2x}^T \epsilon C  \ .
\ee
 In terms of $\psi$ and $\chi$ one obtains the mirror fermion action for
 chiral gauge fields \cite{MIRFER}, which is well suited for
 studying the physically relevant phase with broken symmetry.
\begin{figure}
\vspace{10cm}
\includegraphics{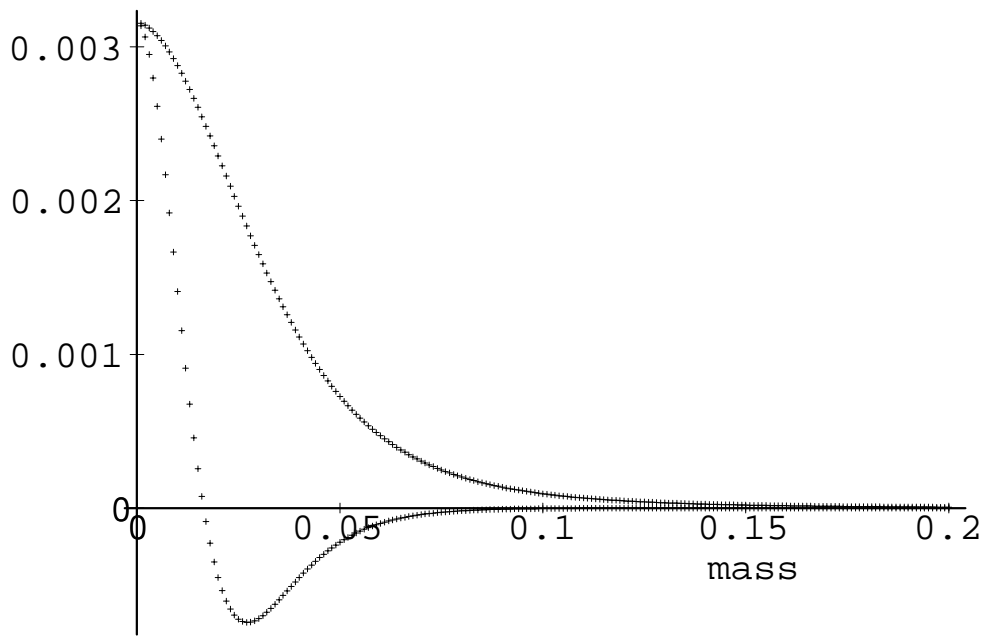}
\caption{ \label{fig1}
The value of the lattice sum ${\cal I_L}$ on $100^4$ lattice
as a function af the bare mass $\mu_0$.
The upper curve is for periodic, the lower one for antiperiodic
boundary conditions.}
\end{figure}
\begin{figure}
\vspace{10cm}
\includegraphics{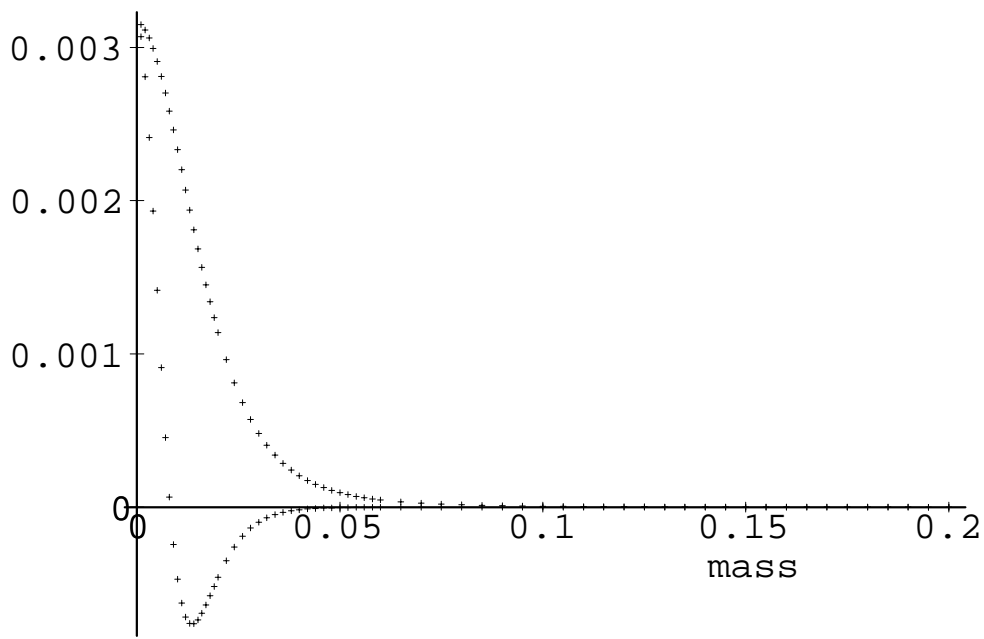}
\caption{ \label{fig2}
The same as fig. 1 on $200^4$ lattice.}
\end{figure}

 In the symmetric (i.~e. confinement) phase there is a natural
 alternative choice in terms of the reshuffled combinations
\be \label{eq15}
\psi_{Cx} \equiv \psi_{Lx}+\chi_{Rx} \ , \hspace{2em}
\psi_{Nx} \equiv \chi_{Lx}+\psi_{Rx} \ .
\ee
 On this basis the vectorlike nature of the model becomes explicit
 ($\gamma_5$'s appear only in the Yukawa-couplings).
 The $\rm SU(2)$ gauge field couples only to the {\em ``charged
 field''} $\psi_C$, and the {\em neutral doublet} $\psi_N$ has only its
 Yukawa-coupling.

 The fermion number is the difference of the number of fermions
 ($\psi$-field) and mirror fermions ($\chi$-field).
 The gauge invariant fermion number current can be defined as
\be \label{eq16}
J_{x\mu} \equiv \frac{1}{2} \left[
(\overline{\psi}_{x+\hat{\mu}} \gamma_\mu U_{Lx\mu} \psi_x)
+(\overline{\psi}_x \gamma_\mu U_{Lx\mu}^\dagger \psi_{x+\hat{\mu}})
-(\overline{\chi}_{x+\hat{\mu}} \gamma_\mu U_{Rx\mu} \chi_x)
-(\overline{\chi}_x \gamma_\mu U_{Rx\mu}^\dagger \chi_{x+\hat{\mu}})
\right] \ .
\ee

\subsection{ Volume dependence }

 The anomalous Ward-Takahashi identity can be derived, as usual, on
 a weak and smooth backgroud gauge field with field strength
\be \label{eq17}
F_{\mu\nu}^s(x) = \partial_\mu A_\nu^s(x) - \partial_\nu A_\mu^s(x)
+ g\epsilon_{stu} A_\mu^t(x) A_\nu^u(x) \ .
\ee
 In the continuum limit of the backgroud field the result for $2N_f$
 fermion doublets is
\be \label{eq18}
\langle \partial_\mu J_\mu(x) \rangle_f =
\lim_{a \to 0} \langle \Delta^b_\mu J_{x\mu} \rangle_f a^{-4}
= N_f g^2\epsilon_{\mu\nu\rho\sigma}
F_{\mu\nu}^s(x)F_{\rho\sigma}^s(x) {\cal I}(r,\mu_0) \ .
\ee
 Here the lattice integral ${\cal I}$ is given by
\be \label{eq19}
{\cal I}(r,\mu_0) \equiv \frac{1}{(2\pi)^4} \int_{-\pi}^\pi
\frac{\mu_k \cos k_1 \cos k_2 \cos k_3 \cos k_4}
{(\bar{k}^2 + \mu_k^2)^3}
[ r \sum_{\alpha=1}^4 \bar{k}_\alpha^2/\cos k_\alpha - \mu_k ] d^4 k\ ,
\ee
 and the notations are
\be \label{eq20}
\mu_k = \mu_0 + \frac{r}{2} \hat{k}^2 \ ,  \hspace{2em}
\bar{k}_\mu = \sin k_\mu \ ,               \hspace{2em}
\hat{k}_\mu = 2\sin \frac{k_\mu}{2} \ .
\ee
 The integral ${\cal I}$ is the same as the one occuring in the chiral
 anomaly, and one can prove (see e.~g. \cite{KARSMI,SEISTA})
\be \label{eq21}
{\cal I}(r,0) = \frac{1}{32\pi^2} \hspace{2em}
({\rm independently\; from}\; r) \ .
\ee

 (\ref{eq18}) and (\ref{eq21}) show that the correct continuum anomaly
 is reproduced at vanishing bare (Majorana-) fermion mass $\mu_0=0$.
 It is, however, interesting to investigate the $\mu_0$ dependence
 of the lattice integral in (\ref{eq19}).
 The numerical evaluation of the corresponding lattice sum ${\cal I}_L$
 on finite ($L^4$) lattices shows that
 ${\cal I} = \lim_{L \to \infty} {\cal I}_L$ is very small,
 probably ${\cal I}(\mu_0,r)=0$ for every positive $\mu_0$.
 (See fig. 1 and 2.)
 The deviation from zero for small $\mu_0$ depends on the boundary
 condition and, as it can be seen, is a function of $L\mu_0$.
 The value of ${\cal I}_L$ comes close to zero near
 $L\mu_0 \simeq 10$.
 This behaviour implies that at every positive $\mu_0$ the
 non-conservation of the current (\ref{eq16}) resulting from the
 topological charge of the background gauge field exactly cancels
 the non-conservation due to the finite bare mass.

 In case of a quantized gauge field, on a fluctuating background,
 the fermion bilinears and the topological density are renormalized
 and mixed with each other.
 The form of the anomaly equation becomes a matter of convention
 (see, for instance, \cite{ESPTAR}).

\section{ Numerical simulations in a 2d U(1) model }

 Before doing numerical simulations in the model for anomalous fermion
 number violation discussed in the previous section, it is useful
 to study a corresponding U(1) toy model in two dimensions, which has
 often been studied in this context (see e.~g. \cite{BOCSHA}).

 The lattice action depending on the compact U(1) gauge field
 $U_{x\mu}=\exp(iA_\mu(x)),\; (\mu=1,2)$ and, for simplicity, fixed
 length Higgs scalar field $\phi(x),\; |\phi(x)|=1$ can be
 written as
\be \label{eq22}
S = \beta \sum_x \sum_{\mu=1, \nu=2}
[1 - cos(F_{\mu\nu}(x))]
-2\kappa \sum_x \sum_{\mu=1}^2 \phi^*(x+\hat{\mu})U_{x\mu}\phi(x) \ ,
\ee
 where the lattice field strength is defined for
 $\mu=1,\nu=2$ as
\be \label{eq23}
F_{\mu\nu}(x) =
A_\nu(x+\hat{\mu})-A_\nu(x)-A_\mu(x+\hat{\nu})+A_\mu(x) \ .
\ee
 Real angular variables $-\pi < \theta_{x\mu} \leq \pi$ on the links
 can be introduced by
\be \label{eq24}
U_{x\mu} \equiv \exp(i\theta_{x\mu}) \ ,
\hspace{2em}
\theta_{x\mu} = A_\mu(x)-2\pi \cdot NINT(A_\mu(x)/2\pi) \ .
\ee
 Fermions in this two dimensional model are introduced in the
 mirror fermion basis $(\psi,\chi)$, according to (\ref{eq14}), but
 will not be explicitly considered here.

\subsection{ Topological charge }

 The topological charge of U(1) lattice gauge field configurations
 can be defined as a sum over the contributions of plaquettes.
 The basic assumption is the existence of a piecewise continuous
 interpolation of the gauge field \cite{LUSCH,GKSW}.
 The gauge invariant topological charge on the torus corresponding to
 periodic boundary conditions is obtained either from the
 ``transition functions'' \cite{LUSCH} or from the ``sections''
 \cite{GKLSW} of this interpolated gauge field.

 Introducing the plaquette angle $-\pi < \theta_{x\mu\nu} \leq \pi$ by
\be \label{eq25}
\theta_{x\mu\nu} \equiv
\theta_{x\mu} + \theta_{x+\hat{\mu},\nu} -
\theta_{x\nu} - \theta_{x+\hat{\nu},\mu} - 2\pi n_{x\mu\nu} \ ,
\ee
 one can show that the topological charge $Q$ is given by
\be \label{eq26}
Q = \sum_x n_{x12} \ .
\ee

\subsection{ Gauge field kinks }

 It follows from the definition in (\ref{eq25}) that the integers
 $n_{x12}$ can have the values $0,\; \pm 1,\; \pm 2$.
 The plaquettes with $n_{x12} \ne 0$ can be imagined to carry
 {\em ``Dirac strings''} or {\em ``gauge field kinks''} \cite{DEGTOU}.
 These local contributions to the topological charge play an
 important r\^ole in the dynamics of lattice U(1) gauge fields and
 has been intensively studied in the literature.
 For recent references in higher dimensions see, for instance,
 \cite{BOMIMP,POYEZU}.
\begin{figure}
\vspace{20cm}
\includegraphics{jhfig03.ps}
\caption{ \label{fig3}
The plaquette contributions to the topological charge $q \equiv n_{12}$
on $64^2$ lattice at $\beta=8.0,\; \kappa=1.0$.}
\end{figure}

 In the scalar-gauge model defined by the lattice action (\ref{eq22}),
 where the pure gauge part corresponds to the Wilson action with
 compact U(1) variables, in a typical configuration there are a lot
 of gauge field kinks.
 As an example, a typical configuration at the large $\beta,\kappa$
 values $\beta=8.0,\; \kappa=1.0$ is shown in fig. 3.
 As one can see, a considerable part of the plaquettes has a kink
 (with positive $n_{12}$) or antikink (negative $n_{12}$).

 It is important to keep in mind that the individual terms in the
 sum over plaquettes (\ref{eq26}) are not gauge invariant, in contrast
 to the total topological charge $Q$.
 Namely, performing large gauge transformations on the two ends of a
 link can create or annihilate a kink-antikink pair on the two
 plaquettes which both contain the link.

 In order to decrease the number of gauge kinks one can fix the gauge
 in some way.
 One possibility is to define the {\em ``minimal gauge''} by
 minimizing the sum of squares (or absolute value) of the link angles
 \cite{GJJLNR}:
\be \label{eq27}
\sum_x \sum_{\mu=1}^2 \theta_{x\mu}^2 \ .
\ee
 This removes some part of the kinks in fig. 3, but still a lot of
 them remains.

 The r\^ole of the gauge kinks can be important for quantitites
 sensitive to the topological charge, or to the smoothness of the
 gauge field.
 Therefore it is instructive to study the model with some modified
 actions suppressing kinks.
 One possibility is to suppress the kinks by introducing a
 ``chemical potential'' for them \cite{BASHSC}, which can in fact also
 completely remove them from the space of allowed configurations.

 Another modification of the gauge field action is suggested by the
 distribution of link angles in the minimal gauge.
 For small gauge couplings $g^2 \equiv \beta^{-1} \ll 1$, where most of
 the link variables are concentrated near $\theta_{x\mu}=0$, there are,
 namely, also some secondary maxima at $\theta_{x\mu}=\pm \pi/2$.
 This is shown on the example of a configuration at
 $\beta=8.0,\;\kappa=0.8$ in fig. 4.
\begin{figure}
\vspace{10cm}
\includegraphics{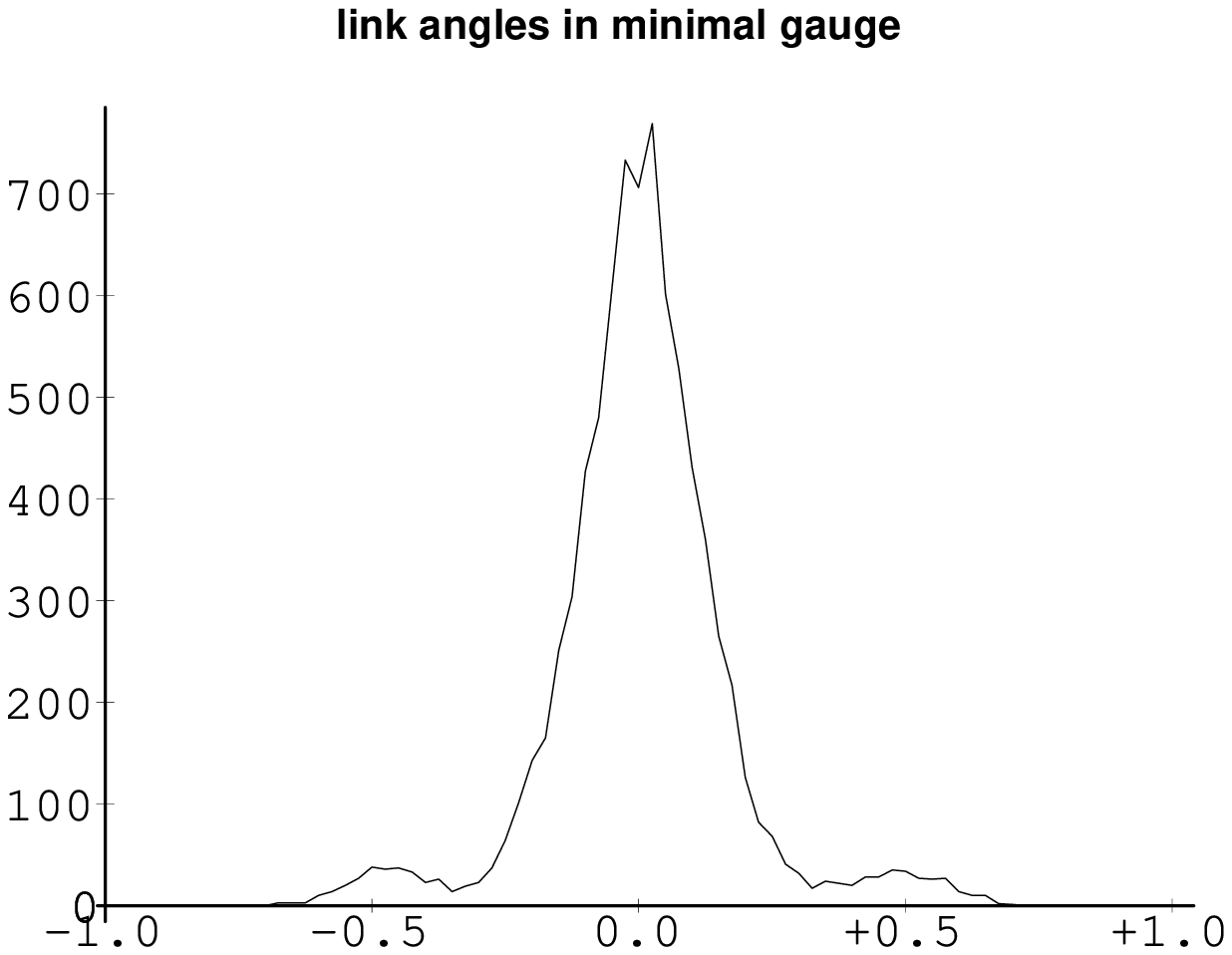}
\caption{ \label{fig4}
The distribution of link angles (in units of $\pi$) of a typical
configuration in the minimal gauge at $\beta=8.0,\; \kappa=0.8$ on
$64^2$ lattice.}
\end{figure}
 The secondary peaks are due to the fact that the Wilson U(1) gauge
 action is periodic for a shift of the plaquette angles by
 $4 \cdot \pi/2 = 2\pi$.
 Therefore, one way to push in the $\beta \to \infty$ limit all link
 angles to zero is to introduce a modified gauge field action like
\be \label{eq28}
S_4 = 16\beta_4 \sum_x \sum_{\mu=1, \nu=2}
\{ 1 - cos[(\theta_{x\mu} + \theta_{x+\hat{\mu},\nu} -
\theta_{x\nu} - \theta_{x+\hat{\nu},\mu})/4] \} \ .
\ee
 The factor 16 in front of $\beta_4$ is introduced in order to have
 in the continuum limit of the action the same normalization as for
 $\beta$.

 The lattice action in (\ref{eq28}) is not exactly gauge invariant
 because large gauge transformations of the links can cause a jump
 by $2\pi$ in the link angles, due to the limitation in (\ref{eq24}).
 This kind of gauge non-invariance is, however, natural since we want
 to suppress exactly such large gauge transformations creating
 kink-antikink pairs.
 $S_4$ remains gauge invariant if both link angles and gauge
 transformation angles are kept small, for instance, less than $\pi/3$
 in absolute value.
 Therefore, the breaking of gauge invariance does not influence
 weak and smooth fields relevant in perturbation theory.

 The modified gauge action $S_4$ suppressing kinks is by no means
 unique.
 One could, for instance, also use the {\em ``non-compact''}
 formulation with the gauge field action
\be \label{eq29}
S_{NC} = \frac{\beta}{2} \sum_x \sum_{\mu=1, \nu=2}
F_{\mu\nu}(x) F_{\mu\nu}(x) \ .
\ee
 We want, however, to stay close to the physically interesting
 model with SU(2) gauge field in four dimensions, where usually
 the compact formulation is used.
 The generalization of the gauge action $S_4$ to SU(2) is possible
 and stays closer to the Wilson action than the non-compact
 formulation.

 The effect of reducing the density of gauge kinks by the modified
 gauge action $S_4$ is dramatic.
 This can be seen on fig. 5.
\begin{figure}
\vspace{20cm}
\includegraphics{jhfig05.ps}
\caption{ \label{fig5}
The same as fig. 3 for the modified gauge action $S_4$ in minimal gauge
at $\beta_4=2.0,\; \kappa=0.8$.}
\end{figure}

\subsection{ Topological susceptibility and C-S numbers }

 An interesting physical quantity is the topological susceptibility
\be \label{eq30}
\chi_Q \equiv \langle Q^2 \rangle - \langle Q \rangle^2 \ .
\ee
 This is strongly influenced by the density of gauge kinks, as is
 shown by figs. 6 and 7.
\begin{figure}
\vspace{10cm}
\includegraphics{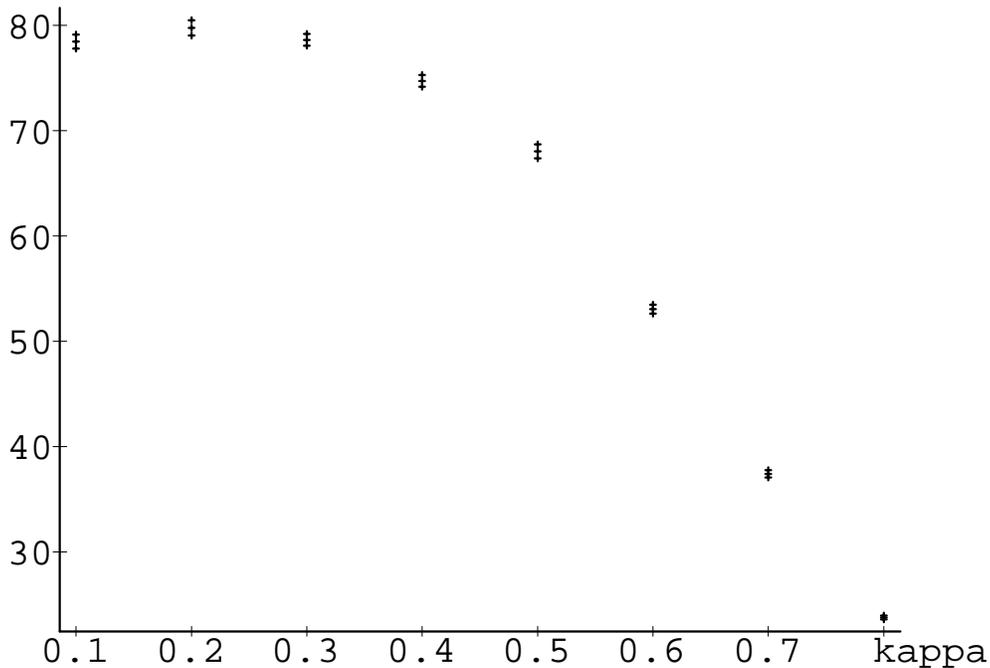}
\caption{ \label{fig6}
The topological susceptibility as function of $\kappa$
at $\beta=2.0$ on $64^2$ lattice.}
\end{figure}
\begin{figure}
\vspace{10cm}
\includegraphics{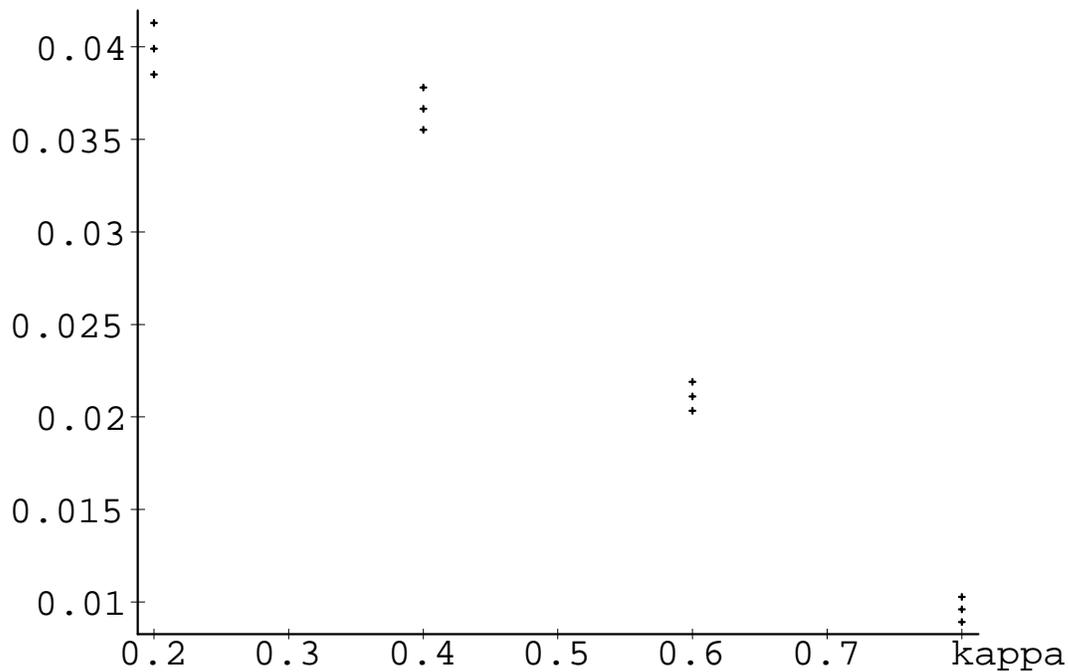}
\caption{ \label{fig7}
The same as fig. 6 with the modified action.}
\end{figure}
 Although the equality of bare couplings $\beta$ and $\beta_4$
 does not mean that, for instance, the correlation lenghts are the
 same, it is clear that $\chi_Q$ gets substantially reduced by the
 suppression of gauge kinks.
 Even if $\beta$ is increased to $\beta=8$, $\chi_Q$ is only reduced by
 roughly a factor of 10, not by a factor more than 1000 as between
 fig. 6 and 7.

 Another important topological feature is the distribution of
 Chern-Simons number.
 This can be defined on a $L_1 \cdot L_2$ lattice in the temporal gauge
 with $U_{x_1,x_2,2}=1,\; x_2=0,1,\ldots,L_2-2$ by the sum of link
 angles
\be \label{eq31}
N_{CS}(x_2) \equiv \frac{1}{2\pi}
\sum_{x_1=0}^{L_1-1} \theta_{x_1,x_2,1} \ .
\ee
 As an example, the distribution is shown in fig. 8 and 9.
\begin{figure}
\vspace{10cm}
\includegraphics{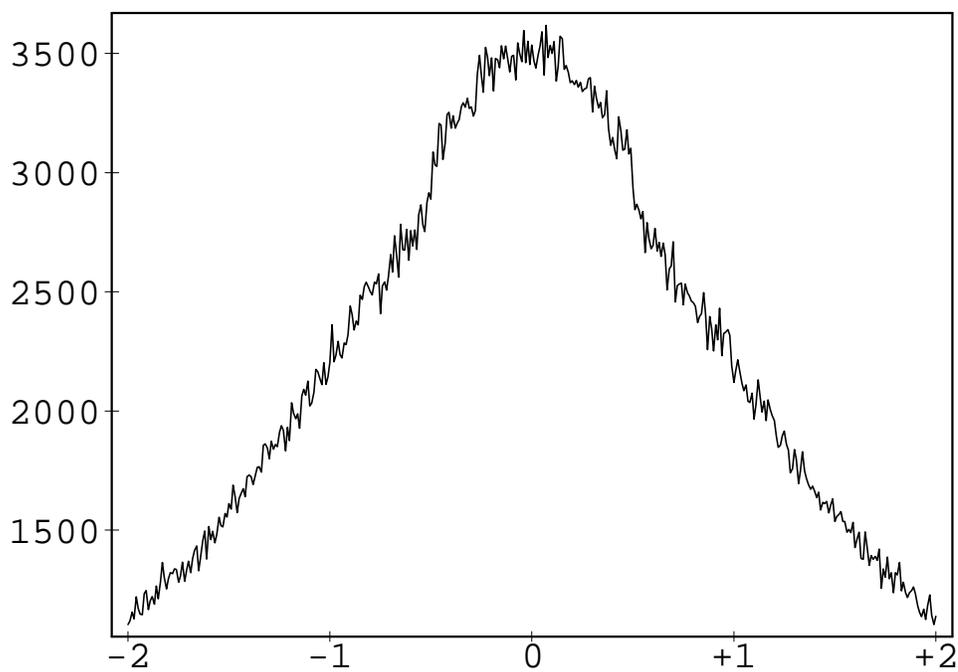}
\caption{ \label{fig8}
The distribution of Chern-Simons number $N_{CS}$ on $64^2$ lattice
at $\beta=8.0,\; \kappa=0.8$.}
\end{figure}

\begin{figure}
\vspace{10cm}
\includegraphics{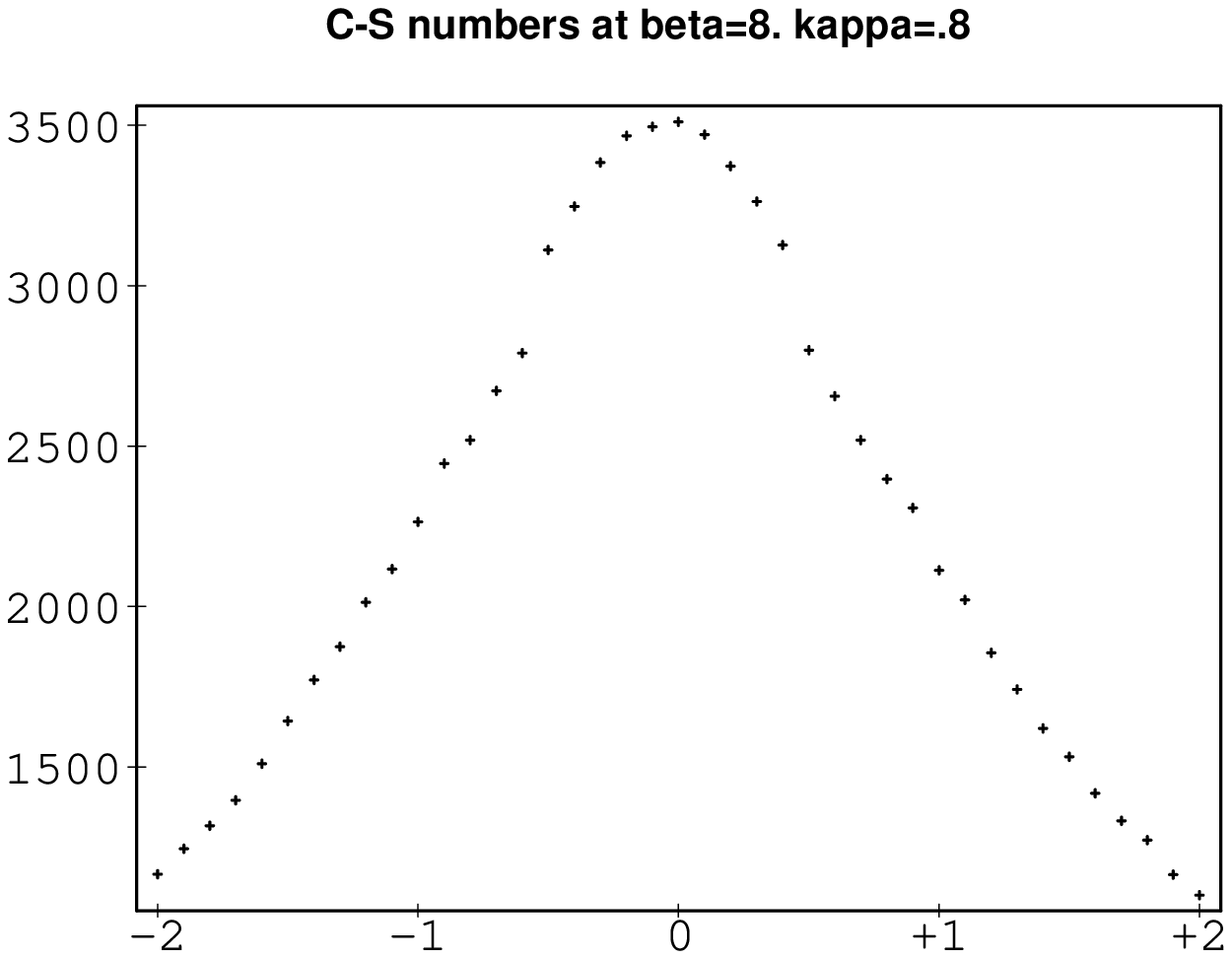}
\caption{ \label{fig9}
The same as fig. 8 in larger bins.}
\end{figure}
 The time dependence (here $x_2$ dependence) of $N_{CS}(x_2)$
 can be characterized by the expectation value of
 $[N_{CS}(x_2)-N_{CS}(x_2-1)]^2$, which is illustrated by fig. 10.
\begin{figure}
\vspace{10cm}
\includegraphics{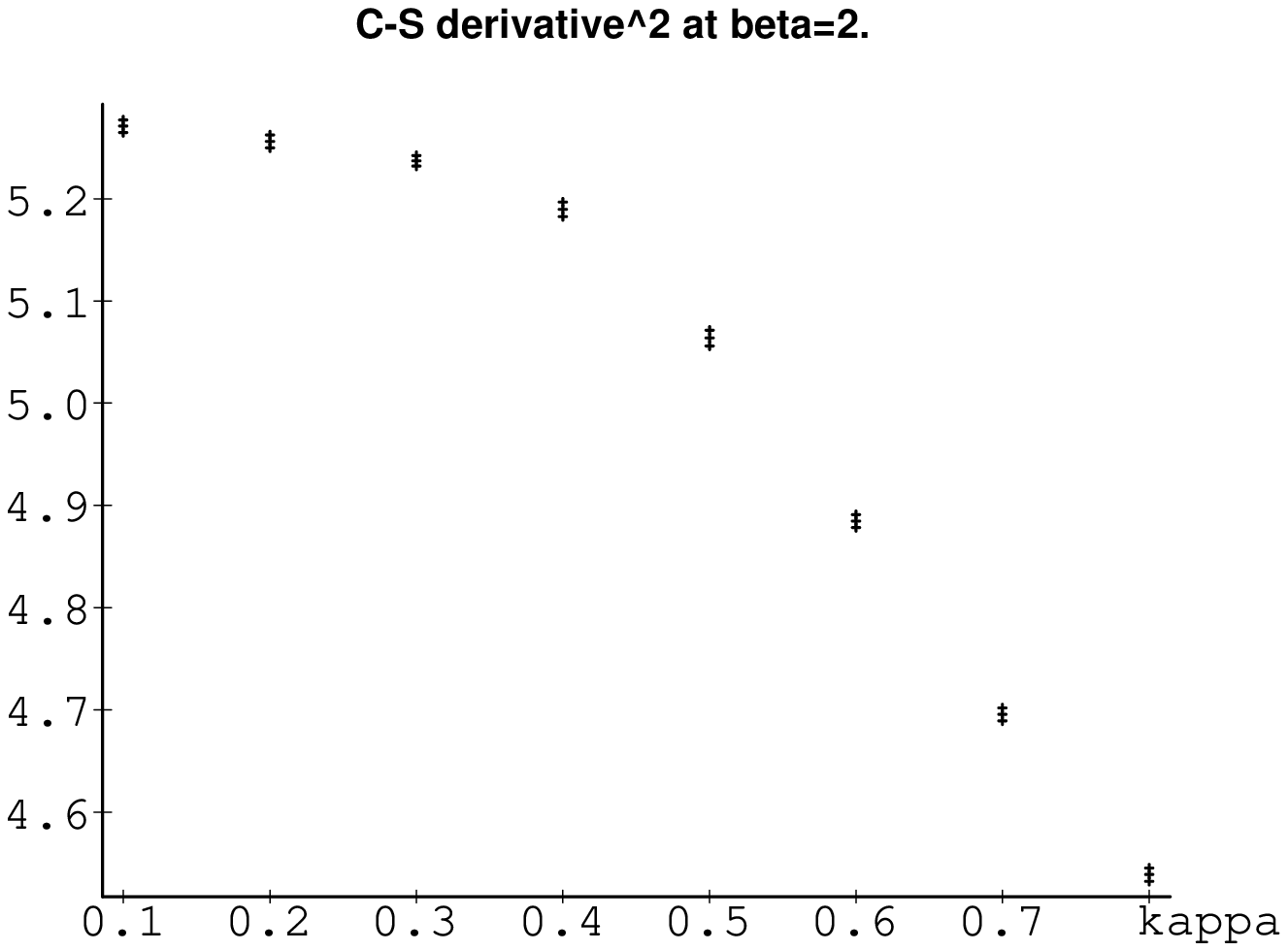}
\caption{ \label{fig10}
The expectation value of the squared time difference of the
Chern-Simons number on $64^2$ lattice
at $\beta=2.0$ as a function of $\kappa$. }
\end{figure}

 In summary, it seems that the density of gauge field kinks has a
 strong influence on the topological features in the two dimensional
 U(1) Higgs model.
 In order to clarify the physical significance of this, detailed studies
 has been started recently \cite{DESMUN}.




\newpage
\begin{thebibliography}{99}
%
\bibitem{THOOFT}
G. 't Hooft,
Phys. Rev. Lett. \underline{37} (1976) 8; \\
Phys. Rev. \underline{D14} (1976) 3432.
%
\bibitem{ROMAPR}
{\it Proceedings of the Rome Workshop on Chiral Gauge Theories,}
Nucl.\ Phys.\  B (Proc.\ Suppl.) \underline{29B,C} (1992).
%
\bibitem{ANFEVI}
I. Montvay,
Nucl. Phys. (Proc. Suppl.) \underline{B30} (1993) 621.
%
\bibitem{CALHAR}
C.G. Callan, J.A. Harvey,
Nucl. Phys. \underline{B250} (1985) 427.
%
\bibitem{KAPLAN}
D.B. Kaplan,
Phys. Lett. \underline{B288} (1992) 342.
%
\bibitem{SHAMIR1}
Y. Shamir,
Weizmann Institute Rehovoth, WIS-93/20/FEB-PH.
%
\bibitem{FUNIEL}
Y. Fu, H.B. Nielsen,
Nucl. Phys. \underline{B236} (1984) 167.
%
\bibitem{KONIPR}
C.P. Korthals-Altes, S. Nicolis, J. Prades,
Marseille, CPT-93/P2920.
%
\bibitem{SHAMIR2}
Y. Shamir,
Weizmann Institute Rehovoth, WIS-93/57/JULY-PH.
%
\bibitem{GRIKIE}
C.J. Griffin, T.D. Kieu,
Melbourne, UM-P-93/05.
%
\bibitem{ROMA2}
A. Borrelli, L. Maiani, G.C. Rossi, R. Sisto, M. Testa,
Nucl. Phys. \underline{B333} (1990) 335;  \\
L. Maiani, G.C. Rossi, M. Testa,
Phys. Lett. \underline{B292} (1992) 397.
%
\bibitem{SMITST}
J. Smit,
Nucl.\ Phys.\  B (Proc.\ Suppl.) \underline{26} (1992) 480; \\
Nucl.\ Phys.\  B (Proc.\ Suppl.) \underline{29B,C} (1992) 83.
%
\bibitem{LEEYAN}
T.D. Lee, C.N. Yang,
Phys. Rev. \underline{104} (1956) 254.
%
\bibitem{MIRFAM}
I. Montvay,
Phys. Lett. \underline{B205} (1988) 315.
%
\bibitem{MIRFER}
I. Montvay,
Phys. Lett. \underline{B199} (1987) 89;    \\
Nucl. Phys. B (Proc. Suppl.) \underline{4} (1988) 443.
%
\bibitem{CSACSI}
C. Cs\'aki, F. Csikor,
Phys. Lett. \underline{B309} (1993) 103.
%
\bibitem{CSIFOD}
F. Csikor, Z. Fodor,
Phys. Lett. \underline{B287} (1992) 358.
%
\bibitem{LANLON}
P. Langacker, D. London,
Phys. Rev. \underline{D38} (1988) 244.
%
\bibitem{CSIMON}
F. Csikor, I. Montvay,
Phys. Lett. \underline{B231} (1990) 503;  \\
F. Csikor,
Z. Phys. \underline{49C} (1991) 129.
%
\bibitem{HERA}
F. Boudjema, F. Csikor, A. Djouadi, J.L. Kneur, I. Montvay, M. Spira,
P.M. Zerwas,
in {\em ''Physics at HERA'',} Proceedings of the HERA Workshop,
eds. W. Buchm\"uller, G. Ingelman, DESY 1992, p. 1094.
%
\bibitem{CSIKOR}
F. Csikor,
private communication.
%
\bibitem{NLC}
A. Djouadi, F. Csikor, I. Montvay,
in {\em ``$e^+e^-$ Collisions at 500 GeV: the Physics Potential'',}
ed. P. Zerwas, DESY 92-123, part B, p. 587.
%
\bibitem{GEOGLA}
H. Georgi, S.L. Glashow,
Phys. Rev. Lett. \underline{32} (1974) 438.
%
\bibitem{GOLPET}
M.F.L. Golterman and D.N. Petcher,
Phys. Lett. \underline{B225} (1989) 159.
%
\bibitem{KARSMI}
L.H. Karsten and J. Smit,
Nucl. Phys. \underline{B183} (1981) 103.
%
\bibitem{SEISTA}
E. Seiler and I.O. Stamatescu,
Phys. Rev. \underline{D25} (1982) 2177.
%
\bibitem{ESPTAR}
D. Espriu and R. Tarrach,
Z. Phys. \underline{C16} (1982) 77.
%
\bibitem{BOCSHA}
A.I. Bochkarev, M.E. Shaposhnikov,
Mod. Phys. Lett. \underline{A2} (1987) 991.
%
\bibitem{LUSCH}
M. L\"uscher,
Comm. Math. Phys. \underline{85} (1982) 39.
%
\bibitem{GKSW}
M. G\"ockeler, A.S. Kronfeld, G. Schierholz, U.-J. Wiese,
J\"ulich, HLRZ 92-34.
%
\bibitem{GKLSW}
A.S. Kronfeld, M.L. Laursen, G. Schierholz, U.-J. Wiese,
Nucl. Phys. \underline{B292} (1987) 330;     \\
M. G\"ockeler, A.S. Kronfeld, M.L. Laursen, G. Schierholz,
U.-J. Wiese,
Nucl. Phys. \underline{B292} (1987) 349.
%
\bibitem{DEGTOU}
T.A. DeGrand, D. Toussaint,
Phys. Rev. \underline{D22} (1980) 2478.
%
\bibitem{BOMIMP}
V.G. Bornyakov, V.K. Mitrjushkin, M. M\"uller-Preussker,
Nucl. Phys. B (Proc. Suppl.) \underline{30} (1993) 587;  \\
V.G. Bornyakov, V.K. Mitrjushkin, M. M\"uller-Preussker,
F. Pahl, HU Berlin-IEP-93/2.
%
\bibitem{POYEZU}
K. Yee,
Nucl. Phys. B (Proc. Suppl.) \underline{30} (1993) 583;  \\
M.I. Polikarpov, K. Yee, M.A. Zubkov,
Baton Rouge, LSU-431-93.
%
\bibitem{GJJLNR}
V. Gr\"osch, K. Jansen, J. Jers\'ak, C.B. Lang, T. Neuhaus, C. Rebbi,
Phys. Lett. \underline{B162} (1985) 171.
%
\bibitem{BASHSC}
J.S. Barber, R.E. Shrock, R. Schrader,
Phys. Lett. \underline{B152} (1985) 221;  \\
J.S. Barber, R.E. Shrock,
Nucl. Phys. \underline{B257} (1985) 515.
%
\bibitem{DESMUN}
DESY--University of M\"unster collaboration, in preparation.
%
\end{thebibliography}
\end{document}